\documentclass[11pt,preprint]{aastex}
\usepackage{graphicx}
\usepackage{amssymb}
\usepackage{amsmath}
\usepackage{bm}
\usepackage{appendix}
\usepackage[T1]{fontenc}
\usepackage[latin9]{inputenc}
\usepackage{listings}

\begin{document}

\title{STATISTICS OF X-RAY POLARIZATION MEASUREMENTS}
\author{C. G. Montgomery\altaffilmark{1}}
\affil{42 Blueberry Lane, Peterborough, NH 03458, USA}
%\email{cgm@physics.utoledo.edu}

\and

\author{ J. H. Swank\altaffilmark{2}}
\affil{Code 662, NASA Goddard Space Flight Center} 
\affil{Greenbelt, MD 20771, USA}
\email{jean.swank@nasa.gov}

\altaffiltext{1} {Professor of Physics and Astronomy (retired), University of Toledo, Toledo, OH, USA.}
\altaffiltext{2} {Emeritus, X-Ray Astrophysics Laboratory}

\begin{abstract}

The polarization of an X-ray beam that produces electrons with velocity components
perpendicular to the beam generates an azimuthal distribution of the ejected electrons. We present methods for simulating and for analyzing the angular dependence of electron detections which enable us to derive simple analytical expressions for useful statistical properties of observable data.  The derivations are verified by simulations. While we confirm the results of previous work on this topic, we provide an extension needed for analytical treatment of the full range of possible polarization amplitudes.

\end{abstract}


\clearpage

\section{INTRODUCTION}

X-ray polarimeters that are based on Compton scattering or the photoelectric effect lead to a distribution of events assigned to 
azimuthal angles in the plane perpendicular to the X-ray beam. The events are detected photons in the case of scattering experiments and detected electrons in the experiments using the photoelectric effect. As in the case of polarization of optical  and longer wavelengths, the polarization of the incoming radiation leads to a distribution  
\begin{equation}
f(\phi)=I_{f}+U_{f}\cos\left(2\phi\right)+Q_{f}\sin\left(2\phi\right),
\end{equation}
where $I_{f}$, $U_{f}$, and $Q_{f}$ describe the distribution of electrons and should 
reflect the Stokes parameters of the incoming X-ray radiation. In general, detector characteristics and background
cause the modulation of the signal to differ from that of the incoming beam. In this paper we address
only the observed modulation. It would be that of the incoming beam for a perfect detector with negligible background. 

Then the polarization amplitude and angle are 
\begin{equation}
a_{e}=\left(U_{f}^{2}+Q_{f}^{2}\right)^{1/2}/I_{f}
\end{equation}
 and the polarization angle estimate is 
\begin{equation}
\phi_{e}=\frac{1}{2}\arctan(Q_{f}/U_{f}).
\end{equation}

Recently the appropriate errors on the determination of X-ray polarization were discussed by \citet[][hereafter Paper 1]{SK13}, \citet{EOW12}, \citet{WEO10}, and \citet{WEK09}. The best estimates and errors on the measurements of optical, infrared, and radio polarization have been discussed in a number of publications taking into account the techniques used for the measurements and their associated experimental errors 
\citep{Vai06, SS85, CSSB83, WK74, Ser62}.
While the polarization amplitude and angle are intuitively the focus of attention, they are not linearly related to the measurables and not independent. It has been pointed out that the Stokes $U_{f}$ and $Q_{f}$ are more amenable to statistical treatment. The same is true for X-ray polarization. 

It has often been assumed that $U_{f}/I_{f}$ and $Q_{f}/I_{f}$  are independent, normally distributed variables. These then determine  the probability of  measured amplitude $a_{e}$ and position angle $\phi_{e}$, given the true amplitude $a_{0}$ and position angle $\phi_{0}$:
\begin{equation}
P(N, a_{e}, a_{0}, \phi_{e}, \phi_{0}) =\frac{Na_{e}}{4\pi}\exp[-\frac{N}{4}(a_{e}^{2} + a_{0}^{2} -2 a_{e} a_{0} \cos2(\phi_{e} - \phi_{0}))]. \label{eq:probability_a}
\end{equation}
Strohmayer \& Kallman (2013)  did extensive simulations for the range of polarization amplitudes $0-1$. While there was agreement for small amplitudes, it was found that the simulation results differed as the amplitude approached $1$. 

We also carried out simulations, with a different approach, but also assuming Poisson errors on the number of events.  We have kept the same convention for $U_{f}$ and $Q_{f}$ as used in Paper I, though other conventions are used, for example in radio astronomy \citep{HB96}.
We likewise see divergence from Equation \ref{eq:probability_a} for the joint probability in angle and amplitude.  It is the purpose of this paper to point out that the divergence we find has a simple description that can be analytically derived. In Section 2 we describe our simulations and the analytical description. In Section 3 we ask what the confidence regions would be for the polarization, given a measurement. Section 4 summarizes the results.

\section{GENERATION AND ANALYSIS OF SIMULATED DATA}

We consider the case of measurements using the photoelectric effect. 
The partially polarized incident beam of X-rays can scatter
electrons into various azimuthal angles around the axis of the incident
beam. We want to estimate the degree of polarization in the incident
beam, and its direction, from the observed numbers of electrons scattered
into different azimuthal angles, and also the uncertainties in those
estimates. 

The polarized component of the beam scatters electrons into an azimuthal
angle $\phi$, measured from the direction of the incident polarization,
with a probability proportional to $\cos^{2}(\phi)$, while
the unpolarized component of the beam scatters electrons isotropically.
In general, if the true polarization is at an angle $\phi_{0}$ in a coordinate system,
we would replace $\phi$ by $(\phi-\phi_{0})$.

So we suppose that the number of scattered electrons reaching detectors,
per unit time, at an azimuthal angle between $\phi$ and $(\phi+d\phi)$, 
can be expressed as proportional to 
$[I_{0}+U_{0}\cos(2\phi)]d\phi$.

 This form ranges from $I_{0}-U_{0}$ to $I_{0}+U_{0}$ and has an
amplitude of variation $a_{0}=U_{0}/I_{0}$. \\
$a_{0}$ is one of the physically significant quantities that we
would like to estimate.

The detectors are taken to divide the interval from $-\pi$ to $+\pi$
into $M$ equal angular bins, each of size $2\pi/M$ radians. $M$
is an odd integer large compared with unity. The bins can be labelled
with an index $j$ which runs from $-(M-1)/2$ to $+(M-1)/2$, with
the center of the $j$th bin at $\phi_{j}=2\pi j/M$.

The expected number of counts in the $j$th bin in a time $T$ is
\begin{equation}
\langle  n_{j} \rangle =\frac{\kappa}{M}[I_{0}+U_{0}\cos(2\phi_{j})]=\frac{\kappa I_{0}}{M}[1+a_{0}\cos(2\phi_{j})].
\end{equation}
$\kappa$ includes geometric and efficiency factors; it is proportional
to $T$.

The values of $n_{j}$ are independent Poisson-distributed random
variables, with these mean values (and variances). The total number
of counts
\begin{equation}
N=\sum_{j}n_{j}
\end{equation}
 is then Poisson-distributed with a mean (and variance) equal to
\begin{equation}
\langle  N \rangle =\kappa I_{0}.
\end{equation}
So a set of samples generated with the same $\langle  N \rangle $ will have different
total count numbers. 

The mathematical properties of the $n_{j}$
values enable deriving a number
of useful facts about the statistical properties of various quantities
related to them.

To estimate the incident polarization amount and direction from
an observed set of $n_{j}$ values, we seek to represent their angular
distribution by a function describing the counts per unit angle.
\begin{equation}
f(\phi)=I_{f}+U_{f}\cos(2\phi)+Q_{f}\sin(2\phi).
\end{equation}

Then the amount and direction of the incoming beam can be estimated
in the usual way. For the amplitude of polarization the estimate is
\begin{equation}
a_{e}=\left(U_{f}^{2}+Q_{f}^{2}\right)^{1/2}/I_{f}
\end{equation}
 and the polarization angle estimate is 
\begin{equation}
\phi_{e}=\frac{1}{2}\arctan(Q_{f}/U_{f}).
\end{equation}
The arctangent function in this equation is actually the two-argument
arctangent function {[}represented in many programming languages as
$\arctan2(Q_{f,},U_{f})${]}. Technically, it is the principal value
of the argument of the complex number $U_{f}+iQ_{f}$. Its values range
from $-\pi$ to $+\pi$ radians, so $\phi_{e}$ ranges from
$-90\degr$ to $+90\degr$. 

Multiplying the function $f(\phi)$ by any constant produces
no change in the amplitude and angle estimates. 

There is a useful graphical representation of the results.
To simplify the notation somewhat, let $U'$ stand for $U_{f}/I_{f}$
and $Q'$ stand for $Q_{f}/I_{f}$. Then the amplitude and angle estimates
are just given by 
\begin{equation}
a_{e}=\sqrt{U'^{2}+Q'^{2}}
\end{equation}
and
\begin{equation}
\phi_{e}=\frac{1}{2}\arctan(Q'/U').
\end{equation}
If we think of $U'$ and $Q'$ as the rectangular coordinates of
a point, these are the polar coordinates of that point: $a_{e}$ is
the distance of the point from the origin and $2\phi_{e}$ is the
angle between the line from the origin to the point and the $U'$
axis. It will be seen that it can be convenient to discuss the geometry
of $U'$ and $Q'$ in the $U'-Q'$ plane and then, if desired, represent
the results in plots of $\phi_{e}$ and $a_{e}$.

Choosing values of $I_{f}$, $U_{f}$, and $Q_{f}$ that best represent
the angular distribution of the set of $n_{j}$ values could be done
in several ways. If the amount of data is large enough, the results
should be essentially the same. One approach would be to find the
values that make the function $f$ the best fit, in some sense, to
the points. This was used in Paper 1, using chi-squared fitting.

Another approach, which is used here, is to view $f$ as part of a
discrete trigonometric interpolating polynomial, where the terms in
the polynomial series are the ones that reflect the properties of
the distribution that are of physical interest. We choose to express
the angular distribution as
\begin{equation}
f(\phi_{j})=\frac{M}{2\pi}n_{j}
\end{equation}
Then $I_{f}$, $U_{f}$, and $Q_{f}$ are the three coefficients that
reflect the polarization of the incoming radiation and that enable
estimates of its direction and amplitude.

The coefficients $I_{f}$ , $U_{f}$ , and $Q_{f}$ are sums over
the angles $\phi_{j}$: 
\begin{equation}
I_{f}=\frac{1}{2\pi}\sum_{j}n_{j}
\end{equation}
\[
U_{f}=\frac{1}{\pi}\sum_{j}n_{j}\cos(2\phi_{j})
\]
\[
Q_{f}=\frac{1}{\pi}\sum_{j}n_{j}\sin(2\phi_{j})
\]
The coefficients $I_{f}$, $U_{f}$, and $Q_{f}$ are random variables,
the sums of linear combinations of the random $n_{j}$ values. From
the Lyapunov central limit theorem their distributions become normal
as $M$ becomes large enough, with means and variances which can be calculated using
the sums over $j$ of polynomials in $\cos(2\phi_{j})$ and $\sin(2\phi_{j})$. $M$ 
does not appear in the limit. Our simulations were insensitive to $M$ above 15.

The mean value of $I_{f}$ is 
\begin{equation}
\langle  I_{f} \rangle =\frac{1}{2\pi}\sum_{j}\langle  n_{j} \rangle =\kappa I_{0}/(2\pi)=\langle  N \rangle /(2\pi)
\end{equation}
The other mean values are
\begin{equation}
\langle  U_{f} \rangle =\kappa U_{0}/(2\pi)=\langle  N \rangle a_{0}/(2\pi)
\end{equation}
\[
\langle  Q_{f} \rangle =0
\]
The variance of $I_{f}$ is 
\begin{equation}
\rm{Var}(I_{f})=\frac{1}{4\pi^{2}}\sum_{j} \rm{Var}(n_{j})=\frac{1}{4\pi^{2}}\sum_{j}\langle  n_{j \rangle }=\langle  N \rangle /(4\pi^{2})
\end{equation}
and by similar calculations the variances of $U_{f}$ and $Q_{f}$
are each equal to
\begin{equation}
\sigma_{f}^{2}=\kappa I_{0}/(2\pi^{2})=\langle  N \rangle /(2\pi^{2}).
\end{equation}
The covariance of $Q_{f}$ with either $I_{f}$ or $U_{f}$ is zero,
but the covariance of $U_{f}$ with $I_{f}$ is $\langle  N \rangle a_{0}/4\pi^{2}$ and the correlation of $U_{f}$ and 
$I_{f}$ is $a_{0}/\surd 2$.
Examples of correlation of $U_{f}$ with $I_{f}$  and $Q_{f}$ with $I_{f}$ were shown in Paper 1. 
For $\phi_{0} \neq 0$,  the correlation of $Q_{f}$ with $I_{f}$ is also nonzero.

The coefficients $I_{f}$, $U_{f}$,  and $Q_{f}$ are not statistically
independent, since they are calculated from the same set of $n_{j}$
values. In fact $I_{f}$,  $U_{f}$, and $Q_{f}$ are trivariate
normal. This is not directly useful since the two quantities of interest
are the ratios $U_{f}/I_{f}$ and $Q_{f}/I_{f}$. Correct distributions
for $U'$ and $Q'$ and thus for $a_{e}$ and $\phi_{e}$ can be generated
by simulations in which the $M$ values of $n_{j}$ are randomly generated
for each sample, and the resulting values of $I_{f}$,  $U_{f}$, 
$Q_{f}$  and thus $U'$ and $Q'$, can be used to get $a_{e}$
and $\phi_{e}$.

The points for a collection of samples will be centered around the $U', Q'$
point with coordinates $a_{0}$, $0$. Let us call that point $Z$.
The probability density for points, that is, the probability per unit
area in this plane, describes what fraction of the points from a set
of many samples will have those locations. A line of constant probability,
a closed curve enclosing the point $Z$, identifies the region within
which a specific fraction of samples will lie, and thus the likelihood
of that set of amplitude and angle estimates. Figure 1 shows an example
for a low true amplitude.

Simplifications result if the
variables $U'$ and $Q'$ are treated as if they were statistically
independent random variables with known means and variances. In fact
the available information allows a calculation of the marginal distribution
of $U'$ and $Q'$ which shows that for large $\langle  N \rangle $ they are independent.
Details are given in Appendix \ref{AppPredicted}, which also provides the joint distribution
of $U'$ and $Q'$ before any approximation of large $\langle  N \rangle $. 

The calculation also provides the means and variances of $U'$ and
$Q'$:
\begin{equation}
\langle  U' \rangle =a_{0}\:;\;\;\langle  Q' \rangle =0\:;\;\;
\end{equation}
\begin{equation}
\rm{Var}(U')=\frac{2}{\langle  N \rangle }(1-a_{0}^{2}/2)\:;\: \rm{Var}(Q')=\frac{2}{\langle  N \rangle }.
\end{equation}
These results are confirmed by the results of simulations for the
full range of amplitude values from 0 to 1. 

The probability density for a system point would be the product of
the probability density for $U'$ and the probability density for
$Q'$ . But these are simply Gaussians with the derived variances
and means. Figure 2 shows the results of simulations for a true polarization
$a_{0}=3/4$, along with the Gaussian distributions with the calculated means
and variances. 

With the definition of  $\lambda$ as the ratio of the standard deviations:
\begin{equation}
\lambda^{2}=\frac{\sigma_{Q'}^{2}}{\sigma_{U'}^{2}}=\frac{1}{1-a_{0}^{2}/2}
\end{equation}
and denoting $\sigma_{Q'}$ simply as $\sigma'$,  the probability
per unit area in the $U'Q'$ plane is also a Gaussian (see Equation(\ref{eq:PU'Q'})):
\begin{align}
P_{U'Q'}(U',Q') &=\frac{\lambda}{2\pi\sigma'^{2}}\exp[-(U'-a_{0})^{2}\lambda^{2}/2\sigma'^{2}]\cdot\exp(-Q'^{2}/2\sigma'^{2}) \\
&=\frac{\lambda}{2\pi\sigma'^{2}}\exp(-\frac{D^{2}}{2\sigma'^{2}}), \notag
\end{align}
with
\begin{equation}
D^{2}=(U'-a_{0})^{2}\lambda^{2} + Q'^{2}.  \label{eq:Dsq}
\end{equation}

It is worth noting that while these results were derived with the
point $Z$ on the $U'$ axis (i.e., with $\phi_{0}=0$) they can be expressed
in a way that is more general. Let us define $2\eta$ as the angular
difference between the location of $Z$ and the location of a data point.
When $Z$ is on the $U'$ axis, $\eta$ is just $\phi_{e}$. But in general  $U'$ in
the above equations is $a_{e}\cos(2\eta)$ while $Q'$ is $a_{e}\sin(2\eta)$.
Then Equation (\ref{eq:Dsq}) becomes
\begin{equation}
D^{2}=\frac{[a_{e}\cos(2\eta)-a_{0}]^{2}}{1-a_{0}^{2}/2} + a_{e}^{2}\sin^{2}(2\eta).   \label{eq:dsqmeas}
\end{equation}
In this form it involves only the distances of $Z$ and the data point
from the origin and the angle between them, and is correct for any
choice of a reference direction.

Lines of constant probability in the $U'-Q'$ plane are ellipses,
centered on the point $Z$, with minor axes along the line from the
origin to the point $Z$, and the ratio of major to minor axes equal
to $\lambda$. An example is shown in Figure 3. 
$D$ is the semi major axis of the ellipse. The area
within an ellipse with a given $D$ is $\pi D^{2}/\lambda$. The total
probability of a point lying outside the ellipse with a given $D$ is
just $\exp(-D^{2}/2\sigma'^{2})$ (as long as $D^{2} \langle  \langle  1$.) See Figure 4 for an example of agreement 
between the probability distribution and a simulation. 

This provides an extension of previous work, covering the full range
of incident amplitudes, even when the correlation of $I_{f}$ with
$U_{f}$ changes the variance of $U'$ enough to be important. The results for $U'$ and $Q'$ also give 
\begin{align}
\rm{Var}(a_{e})&= \sigma'^{2} (1-a_{0}^{2}/2), \\
\rm{Var}(\eta)&= \frac{1}{4} \frac{\sigma'^{2}}{a_{0}^{2}}.
\end{align}

Since $\lambda$ is quite close to unity unless the incident polarization
amplitude is quite large, however, a simpler approximation should
often be applicable. Setting $\lambda=1$ reduces $D^{2}$ to 
\begin{equation}
R^{2}=(U'-a_{0})^{2}+Q'^{2}
\end{equation}
This is simply the square of the distance from the point with rectangular
coordinates $U'$ and $Q'$ to the point $Z$. The probability density
is
\begin{equation}
P_{U'Q'}(U',Q')=\frac{1}{2\pi\sigma'^{2}}\exp(-R^{2}/2\sigma'^{2}).
\end{equation}
 The lines of constant probability are simply circles centered at
$Z$. The probability that a point is farther from $Z$ than $R$
is $\exp(-R^{2}/2\sigma'^{2})$. 

It is straightforward to reexpress this in terms of the polar coordinates
$a_{e}$ and $2\phi_{e}$ , noting that the element of area in these
coordinates is $a_{e}\cdot d(a_{e})\cdot d(2\phi_{e})$ . Also, the
variance $\sigma'^{2}=2/\langle  N \rangle $ can be replaced by $2/N$ if the number
of counts is large. The result is 
\begin{equation}
P(a_{e},\phi_{e})=\frac{Na_{e}}{4\pi}\exp[-\frac{N}{4}(a_{e}^{2}+a_{0}^{2}-2a_{e}a_{0}\cos(2\phi_{e}))]
\end{equation}
which is the formula commonly used in discussions of this topic \citep{WEO10, SK13}, the same 
as Equation(\ref{eq:probability_a}) with $\phi_{0}=0$.

As pointed out by those authors, if the angle is not of interest, but only the amplitude of the polarization, the integral over angles gives 
the Rice distribution \citep{R45} for the amplitude, a distribution used in various other signal processing applications. If, however, $a_{0}$
is not small and Equation (\ref{eq:dsqmeas}) is applicable, the angular distribution is in the category of generalizations of the von Mises distribution \citep{MJ99,YB82}. It has been used in studies of Brownian motion, waves, and bending of biological molecules, for examples. The integral over angle gives a more complex distribution in $a_{e}$. This can certainly be numerically computed, if a single parameter confidence region is desired. In this paper we are concentrating on joint confidence regions in the polarization amplitude and direction. 

Comparison to the work of \citet[][Paper 1]{SK13} finds some differences 
in the variances of $U_{f}$ and $Q_{f}$   and their correlations with $I_{f}$ for the variables
calculated in different ways. For a given simulation the results of calculating
$U_{f}$ and $Q_{f}$ from trigonometric interpolating polynomials agree with the values obtained
from unweighted least squares fitting (in accordance with Parseval's theorem), but differ from the results of
least squares weighted inversely with the counts in the bins (which exceeded 100 for all bins in our simulations). 
For the unweighted least squares, the variances of $U_{f}$ and $Q_{f}$ 
are independent of polarization and the correlations strictly linear, while for the weighted least squares, the variances
decrease with amplitude. This in turn causes an amplitude dependent increase in the correlation magnitudes. 
It does not seem surprising that there is a difference in the distribution of the best fits and the shapes of the probability surfaces. 
The mean values of the fits are consistent. 

Some details of our simulations and calculations are given in Appendix \ref{AppComp}.

\section{DRAWING CONCLUSIONS FROM AN OBSERVED DATASET}

We turn now to the question of how we can draw conclusions about the
magnitude and direction of incoming X-rays from a single data set.
We have the number of counts at each azimuthal angle with respect to 
some angle on the sky. We calculate
$I_{f}$, $U_{f}$, and $Q_{f}$ for our angular distribution
of counts, and from them we compute two quantities: $U'$ and $Q'$. We
can think of these as the rectangular components of a point in a coordinate
system with the $U'$ axis in our reference direction. The origin of
this coordinate system is the point corresponding to a completely
unpolarized beam, for which both components would vanish.

What we do not know but would like to draw conclusions about is the
location of the point $Z$ in this coordinate system. By definition,
$Z$ is the point whose distance from the origin is the true polarization
amplitude $a_{0}$ and which is in the same direction as the true
polarization direction.

A contour line enclosing the observed data point
can be defined by the fact that the probability of the observed $U'$ and $Q'$
values for any $Z_{C}$ (the $C$ stands for ``candidate'') located at a point 
on that contour has the same value.
Any point inside the contour is a location for $Z_{C}$ that gives
a probability of the data that is higher, and any point outside the
contour is a location for $Z_{C}$  such that the data has a lower probability.
This defines a confidence region for $Z_{C}$.

We want to find the coordinates of the points $Z_{C}$ for which the probability of the
data point, with coordinates $U'$ and $Q'$, has a specified probability.
The geometry of this situation is very similar to the case
previously studied, where one wanted to find which data points have
a specified probability when the true amplitude is known. We define
the angle $2\eta_{C}$ as the angle between lines from the origin
to the data point and to $Z_{C}$, the difference between the angular
position of the data point and the angular position of the point $Z_{C}$.

In order to have a specified probability, the data point must lie
somewhere on an ellipse whose center is at the point $Z_{C}$, whose
major axis is perpendicular to the line from the origin to $Z_{C}$
and has the length $2D$, and whose minor axis is $2D/\lambda_{C}$,
with $\lambda_{C}=1/\sqrt{1-a_{C}^{2}/2}$. $D$ is related to the
probability in the same way as in the previous discussions; it depends
only on the total number of counts in the data set and the specified
probability.

From the earlier section, the requirement for the desired probability
is given by Equation (\ref{eq:dsqmeas}), which can be rewritten for the present purpose
as 
\begin{equation}
D^{2}=\frac{[a_{e}\cos(2\eta_{C})-a_{C}]^{2}}{1-a_{C}^{2}/2} + a_{e}^{2}\sin^{2}(2\eta_{C}).
\end{equation}

We define a pair of new variables
\begin{equation}
u=a_{C}\cos(2\eta_{C})\:\:;\:\: v=a_{C}\sin(2\eta_{C})
\end{equation}
which specify the location of $Z_{C}$ relative to the data point.
Using these, the condition that the data point have the correct position
relative to $Z_{C}$ can be written as
\begin{equation}
u^{2}(1+D^{2}/2)-2a_{e}u+v^{2}[1+(D^{2}-a_{e}^{2})/2]=D^{2}-a_{e}^{2}   \label{eq:cand_ellipse}
\end{equation}
 This is a quadratic in $u$ and $v$ and thus defines an ellipse.
It is even in $v$, so the axes lie along and perpendicular to the
line from the origin to the data point. The quadratic equation for
$u$ when $v=0$ has two roots equidistant from $u=a_{e}/(1+D^{2}/2)$,
so the center of the ellipse is located at $u=a_{e}/(1+D^{2}/2)\:;\: v=0$.
 The $v$ semi-axis is $D/\sqrt{1+D^{2}/2}$
and the $u$ semi-axis is $[D/(1+D^{2}/2)]\sqrt{1+D^{2}/2-a_{e}^{2}/2}$.
(A more thorough and detailed derivation is provided in Appendix \ref{AppInverse}.)

Since the derivation of probabilities in the earlier section included
neglecting $D^{2}$ compared with unity, it is not inconsistent to
do the same here. The result is simpler than might have been expected.
The ellipse is centered on the data point and has axes of $2D$ and
$2D/\lambda_{e}$. The data point and the candidate have changed places.
Contour lines around the data point (in the $U'Q'$ plane) with a
specified value of $D$ are given by:
\begin{equation}
D^{2}=(u-a_{e})^{2}\lambda_{e}^{2}+v^{2}
\end{equation}
 where $u=a_{C}\cos(2\eta)$ and $v=a_{C}\sin(2\eta)$ as defined
in Equation (31), and $2\eta$ is the angle between the data point and the
point on the contour.

One can then derive a description of probabilities similar to the
situation of a known incident amplitude. The area of an ellipse with
a given value of $D$ would be just $\pi D^{2}/\lambda$, and the
incremental area between ellipses for $D$ and $D+dD$ would be proportional
to $DdD$. An integral from $D$ to infinity is then proportional
to $\exp(-D^{2}/2\sigma'^{2})$. Normalization makes the probability
outside the contour equal to $\exp(-D^{2}/2\sigma'^{2})$.

Figure 5 shows the implication of the results for a particular estimated polarization from a measurement
sample. The marginal probability of $U'$ and $Q'$ derived in Section 1 is in terms of $\langle  N \rangle $ for many samples. 
For a single sample, one has only the $N$ for that sample. The probability depends on $\langle N \rangle D^{2}$ 
and is only significant for small $D^{2}$. The fractional error in $D^{2}$ should be $1/\sqrt{N}$.

It seems likely that in any practical situation there will be large
enough additional uncertainties in data acquisition not included in
this statistical analysis that the difference in the contours for
approximate or full inclusion of $D^{2}$ dependence would not be
important.

It has been mentioned that statistical variations in the recorded
counts result in a positive probability for any pair of $U'$ and
$Q'$ values, even improbable ones which might lead to amplitude estimates
that are larger than 1, while the true amplitude can only be zero
or positive and less than or equal to 1. This could complicate the
correct conclusions that can be drawn from a single data set if the
total number of counts is too low or the confidence level chosen leads
to a formal contour that extends into regions where $U'^{2}+Q'^{2}$ is
greater than 1. The simplest way to avoid such problems is just to
get more data, or choose a different confidence level, or both, so
that the contour is completely within the acceptable range. 
But if this is not feasible or desirable,
one could impose restrictions and provide a truncated confidence region.

\section{SUMMARY}

The methods presented here for generating simulated counts of scattered
electrons at different azimuthal angles, and for analyzing the resulting
angular distribution, have the special advantage of making it possible
to derive rigorous results about the estimation of polarization amplitudes
and directions and their uncertainties. Alternative methods for generating
simulations, and for analyzing the angular distribution, should give
results that are essentially identical to these, but some of the
predicted properties may only be observed rather than derived.

We confirm previous work on amplitude and angle estimates
for cases in which the incident amplitude is not too large. With our approach
we are able to provide analytical treatment for larger incident
amplitudes, so that the entire physical range of possibilities is
covered.

\acknowledgments

We thank Tod Strohmayer, Tim Kallman, and Phil Kaaret for detailed discussions.

\clearpage

\begin{appendices}
\appendix

\numberwithin{equation}{section}
\section{THE DISTRIBUTIONS OF $U_{f}, Q_{f}$ AND OF $U', Q'$ \label{AppPredicted}}

It is useful to introduce another variable, linear in $U_{f}$ and $I_{f}$. Because $\langle  U_{f} \rangle   = a_{0} \langle  I_{f} \rangle  $, the mean value of $(U_{f}-a_{0} I_{f})$ is zero and because the covariance of $I_{f}$ with $U{_f}$ is just $a_{0}$ times the variance of $I_{f}$, the covariance of $(U_{f}-a_{0} I_{f})$ with $I_{f}$ is zero.  The variance of $(U_{f}-a_{0} I_{f})$ is 
$\sigma_{f}^{2} (1-a_{0}^{2}/2)$. We introduce the quantity 
\begin{equation} 
\lambda=1/\sqrt{1-a_{0}^{2}/2}
\end{equation}
 and define the new variable
\begin{equation}
V_{f}=\lambda(U_{f}-a_{0}I_{f}).
\end{equation}
$V_{f}$ has a mean value of zero and the same variance as $Q_{f}$ and is uncorrelated with $I_{f}$ and $Q_{f}$. 

The three variables, $I_{f}$, $V_{f}$, and $Q{_f}$ are trivariate normal, and mutually uncorrelated, so they are independent. 
Then the joint probability distribution function is given by 
\begin{equation}
P_{I,V,Q} ( I_{f}, V_{f}, Q_{f}) = \frac{\sqrt{2}}{(2\pi\sigma_{f}^{2})^{3/2}}\exp[-\frac{1}{2\sigma_{f}^{2}}(2(I_{f}-\langle  I_{f} \rangle  )^{2}+V_{f}^{2}+Q_{f}^{2})]. \label{eq:joint}
\end{equation}

What we would like to have is the joint probability of $Q' = Q_{f}/I_{f}$ and $V' = V_{f}/I_{f} = \lambda (U'- a_{0})$. To obtain this, we transform to $V'$ and $Q'$ as our variables, and integrate the probability of $I_{f}, V', Q'$ over $I_{f}$. For the transformation we have $dV'dQ' = I_{f}^{2} dV_{f} dQ_{f}$. 
Since $V'$ and $Q'$ only appear in the combination 
\begin{equation}
D^{2}= V'^{2} + Q'^{2}
\end{equation} 
and $\langle  I_{f} \rangle   = \langle  N \rangle  /(2\pi) = \pi \sigma^{2}_{f}$, the integral to be evaluated is 
\begin{equation}
P_{V'Q'}(V', Q') = \frac{\sqrt{2}}{(2\pi\sigma^{2}_{f})^{3/2}} \int_{-\infty}^{\infty} dI_{f} I_{f}^{2} \exp[-\frac{I_{f}^{2}}{\sigma_{f}^{2}} (1+D^{2}/2)+2 \pi I_{f} - \pi^{2} \sigma{_f}^{2}],
\end{equation}
which can be evaluated exactly. It is convenient to introduce 
\begin{equation}
\sigma'^{2} = \frac{1}{\pi^{2} \sigma_{f}^{2}} = \frac{2}{\langle  N \rangle  }
\end{equation}. 
Then 
\begin{equation} 
P_{V'Q'}(V',Q') = \frac{1+(1+D^{2}/2)/\langle  N \rangle  }{(1+D^{2}/2)^{5/2}} \frac{1}{2\pi\sigma'^{2}} \exp(-\frac{D^{2}}{2 \sigma'^{2}}). \label{eq:PV'Q'exact}
\end{equation}
In any practical application, the value of $\langle  N \rangle  $ will be so large that the term proportional to $1/\langle  N \rangle  $ can be neglected in comparison with unity. Moreover, the values of $D$ that might be of interest will be small enough to justify replacing $(1+D^{2}/2)$ with unity. Otherwise, the probabilities involved are so small that they would be of no value. Accordingly the initial fraction in Equation(\ref{eq:PV'Q'exact}) can be dropped, leaving
\begin{equation}
P_{V'Q'}(V',Q') = \frac{1}{2 \pi \sigma'^{2}} \exp(-\frac{D^{2}}{2 \sigma'^{2}}).  \label{eq:PV'Q'} 
\end{equation}
The distribution of $D^{2}$ is just a chi-squared distribution with two degrees of freedom.

Thus it is clear that 
\begin{equation}
P_{U'Q'}(U',Q') = \frac{\lambda}{2\pi\sigma'^{2}}\exp(-\frac{D^{2}}{2 \sigma'^{2}}). \label{eq:PU'Q'}
\end{equation}
This is the form in which $U'$ and $Q'$ are approximately independent normal variables with the variance in $U'$ lower than that of $Q'$ by $1-a_{0}^{2}/2$. These results for the variance of $U'$ and $Q'$ agree with the estimates given by considering $\langle  \delta^{2} U' \rangle   $ and $\langle  \delta^{2} Q' \rangle   $, where expanding around the mean values gives $\delta U' = \delta U_{f}/\langle  I_{f} \rangle    -\langle  U_{f} \rangle   \delta I_{f}/\langle  I_{f} \rangle  ^{2}$  and similarly for $\delta Q'$. 

If $\phi_{0} \neq 0$, the following means and covariances have a $\phi_{0}$ dependence:
\begin{align} 
\langle  U_{f} \rangle   &=\frac{\langle  N \rangle  }{2 \pi} a_{0} \cos (2\phi_{0}) , \\
\langle  Q_{f} \rangle   &=\frac{\langle  N \rangle  }{2 \pi} a_{0} \sin (2\phi_{0}) ,   \notag \\
\langle  \delta U_{f} \delta I_{f} \rangle   &= \frac{\langle  N \rangle  }{4 \pi^{2}} a_{0} \cos (2\phi_{0}), \notag \\
\langle  \delta Q_{f} \delta I_{f} \rangle   &= \frac{\langle  N \rangle  }{4 \pi^{2}} a_{0} \sin (2\phi_{0}).   \notag 
\end{align}
It remains true that 
\begin{equation}
\langle  \delta U_{f} \delta Q_{f} \rangle   = 0.                               \notag
\end{equation}
With 
\begin{align}
X &=U \cos (2 \phi_{0})+ Q \sin (2 \phi_{0}), \\
Y &=-U \sin (2 \phi_{0})  + Q \cos   (2 \phi_{0}) ,      \notag     \\            
Z&=\lambda(X-a_{0} I_{f}), \notag
\end{align} 
corresponding to Equation (\ref{eq:joint}), we have
				
\begin{equation}
P_{IZY}(I_{f}, Z_{f}, Y_{f}) =  \frac{\sqrt{2}}{(2\pi\sigma_{f}^{2})^{3/2}}\exp[-\frac{1}{2\sigma_{f}^{2}}(2(I_{f}-\langle  I_{f} \rangle  )^{2}+Z_{f}^{2}+Y_{f}^{2})].
\end{equation}

$X,  Y$ are just a set of axes rotated from $U, Q$ by $2 \phi_{0}$.    Since
\begin{align} 
U &= a_{e} \cos (2 \phi_{e}) \notag \\
Q &= a_{e} \sin (2\phi_{e}), \notag
\end{align}
we have
\begin{align} 
X&=a_{e}\cos (2(\phi_{e} - \phi_{0})) \notag \\
Y&=a_{e} \sin (2(\phi_{e}-\phi_{0})). \notag
\end{align}
The probability only depends on the angle between $ \bm{ a_{e}}$ and the true polarization $ \bm{a_{0}}$, which  is $2 \eta = 2 (\phi_{e}-\phi_{0})$.

With $X'=X/I'$, $Y'=Y/I'$, and $D^{2} = \lambda ^{2} (X' -a_{0})^{2} + Y'^{2}$   
the general result is derived as before, and similarly, for large $\langle  N \rangle  $,
\begin{equation}
P_{X'Y'}(X',Y')   \approx \frac{\lambda}{2\pi \sigma'^{2}} \exp(-\frac{D^{2}}{2 \sigma'^{2}}). 
\end{equation}

\section{COMPUTATIONAL REMARKS \label{AppComp}}

Numerical implementation of the methods presented here for generating and analyzing simulations was done primarily with a simple Fortran program (available from the authors,) which can produce results for thousands of samples with thousands of counts each, in a few seconds on modest hardware.

Other simulations were done using IDL with version 8.2  RANDOMU to generate Poisson variates, both for $n_{j}$ directly for 40 bins and for $N$ in order to use the transformation method used by \citet[][Paper 1]{SK13}. To generate the $n_{j}$ in this case the method of Marsaglia was used \citep{Mars64}. These simulations were carried out to establish that the difference between the $U$ and $Q$ variances was due to the difference between the unweighted and weighted least-squares fits. The fits were carried out using IDL and the fitting routines based on MINPACK-1 \citep{CM09}.

The locus of points $a_{e}$ and $\eta$ that satisfy Equation (\ref{eq:dsqmeas}) can be obtained parametrically. Defining $ \bm{ \rho}$ by 
\boldmath
\begin{equation} 
 a_{e}= a_{0}+ \rho
\end{equation}
\unboldmath
and $\beta$ as the angle between  $ \bm{\rho}$ and $\bm{ a_{0}}$,
$a_{e} \cos (2\chi) = a_{0} + \rho \cos (\beta)$ and
$a_{e} \sin (2 \chi) = \rho \sin (\beta)$. Then Equation (\ref{eq:dsqmeas}) gives
$\rho = D/\sqrt{\sin (\beta^{2}) + \cos (\beta ^{2})/(1-a_{0}^{2}/2)}$
for $ 0 \leq \beta \leq 2 \pi$.
We then calculate 
\begin{align}
a_{e}^{2} &= a_{0}^{2} + \rho^{2} + 2 a_{0} \rho \cos (\beta) ,  \\
\cos (2 \chi) &= \frac{(a_{0} + \rho \cos (\beta))}{a_{e}} ,  \notag \\
\sin (2 \chi) &= \frac{ \rho \sin (\beta)}{a_{e}} , \notag \\
2 \chi &= \arctan 2 (\cos (2 \chi), \sin (2 \chi)). \notag
\end{align}

A similar parametric construction gives the contours of candidate true values of amplitude and angle, given measured values, as discussed in Section 3.

\numberwithin{equation}{section}
\section{DERIVATION OF CANDIDATE CONTOURS \label{AppInverse}}
The  probability of a particular measured polarization, given a known true polarization, depends only on $D^{2}$ (Equation (\ref{eq:dsqmeas}) in the text), where
\begin{equation}
D^{2} =  \frac{[a_{e} \cos(2\eta)-a_{0}]^{2} }{1-a_{0}^{2}/2} + a_{e}^{2} \sin^{2} (2\eta) .
\end{equation}     
We wish to determine the candidate true polarizations $a_{C}$ at angle $2\eta$ from a measured polarization amplitude and direction 
that would have a particular probability. The values must obey the same equation with $a_{0}  \rightarrow a_{C}$.  The candidate point $Z_{C}$ has the projections $u=a_{C} \cos2\eta$  and $v=a_{C} \sin2\eta$ parallel and perpendicular to ${\bf a_{e}}$.
Then, using these substitutions and $a_{C}^{2} =u^{2} + v^{2} $,  Equation (\ref{eq:cand_ellipse}) in the text is obtained, which can be written as
 \begin{equation}
 D^{2} -a_{e}^{2} - (u^{2} + v^{2}) (1+D^{2}/2) + a_{e} ^{2} \frac{v^{2}}{2} + 2 a_{e} u = 0.
 \end{equation}
The following sequence of reorganizations:
\begin{equation}
\frac{(D^{2} - a_{e}^{2})}{1+D^{2}/2} -u^{2} -v^{2} + \frac{a_{e}^{2} v^{2} /2}{1+D^{2} /2} + \frac{2 a_{e} u}{1+D^{2} /2} = 0,
\end{equation}

\begin{equation}
\frac{(D^{2} - a_{e}^{2})}{1+D^{2}/2} -(u -\frac{a_{e} }{1+D^{2} /2} )^{2} + \frac{a_{e}^{2}}{(1+D^{2}/2)^{2}} -v^{2} (1- \frac{a_{e}^{2} /2}{1+D^{2} /2})=0  ,
\end{equation}
\begin{equation}
\frac{D^{2}}{1+D^{2}/2} -\frac{a_{e}^{2}}{1+D^{2}/2}(1-\frac{1}{1+D^{2}/2})-(u -\frac{a_{e} }{1+D^{2} /2} )^{2} -v^{2} (1- \frac{a_{e}^{2} /2}{1+D^{2} /2})=0,
\end{equation}
\begin{equation}
\frac{D^{2}}{1+D^{2}/2}(1 -\frac{a_{e}^{2}/2}{1+D^{2}/2})-(u -\frac{a_{e} }{1+D^{2} /2} )^{2}  -v^{2} (1- \frac{a_{e}^{2} /2}{1+D^{2} /2})=0.
\end{equation}
leads to
\begin{equation}
\frac{D^{2}}{1+D^{2}/2} = (u -\frac{a_{e} }{1+D^{2} /2} )^{2}/ (1- \frac{a_{e}^{2} /2}{1+D^{2} /2}) + v^{2}.
\end{equation}
This is the ellipse centered on $a_{e} /(1+D^{2} /2), 0$ with semi-axes  $ (D/(1+D^{2}/2))\sqrt{1-a^{2}/2+ D^{2}/2}$ and $D/\sqrt{1+D^{2}/2}$
for $u$ and $v$, respectively.
Now neglecting $D^{2}/2$ compared to $1$,
\begin{equation}
D^{2} = \frac{(u -a_{e} )^{2}}{1- a_{e}^{2}/2} + v^{2}.
\end{equation}

\end{appendices}

\clearpage
\begin{figure}
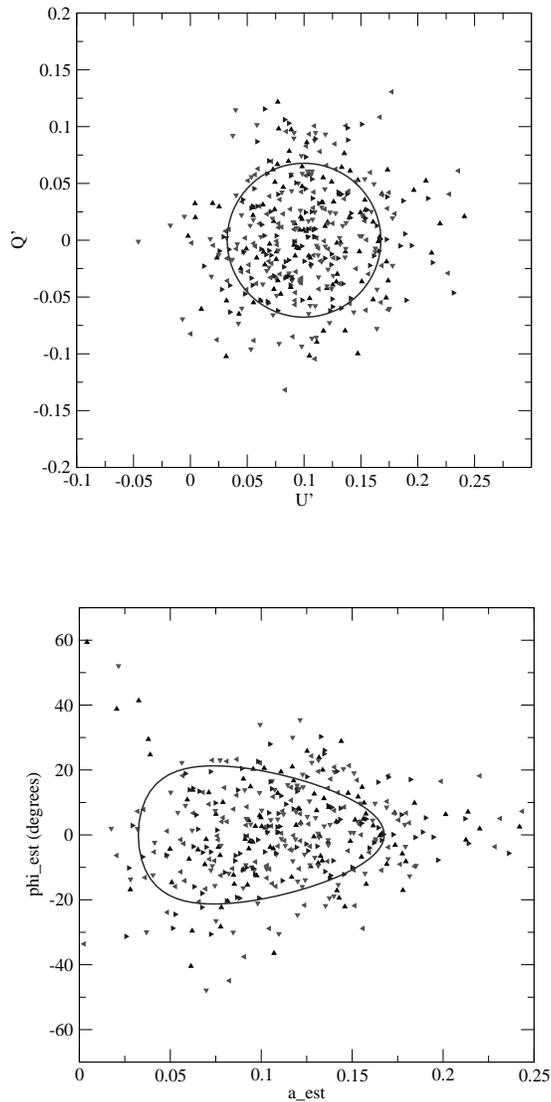


\vspace {-.4in}
\centerline{\includegraphics[angle=0,scale=0.4]{fig1topUQprimeLowPol.eps}}
\vspace{0.5 in}
\hspace{0.05 in}
\centerline{\includegraphics[angle=0,scale=0.4]{fig1bottomaphiLowPol.eps}}
\vspace{-0.2in}
\caption{(top) $U'$ and $Q'$ for 4 sets of 100 simulations each for $<N>=1000$, with the  true polarization along the $U_{f}$ axis. Points from the different sets are indicated by triangles of different orientation. The 68.3 \% probability contour is centered on the point $Z$ at $a_{0} = 0.1, 0$. (bottom) The polarization amplitude $a_{e}$  and angle $\phi_{e}$  corresponding to the $U'$ and $Q'$ results. $\phi_{e}$
is half the angle between the $U'$ axis and the direction to the point $U', Q'$. \label{Figa} }
\end{figure}
\clearpage
\vspace{-.6in}
\begin{figure}
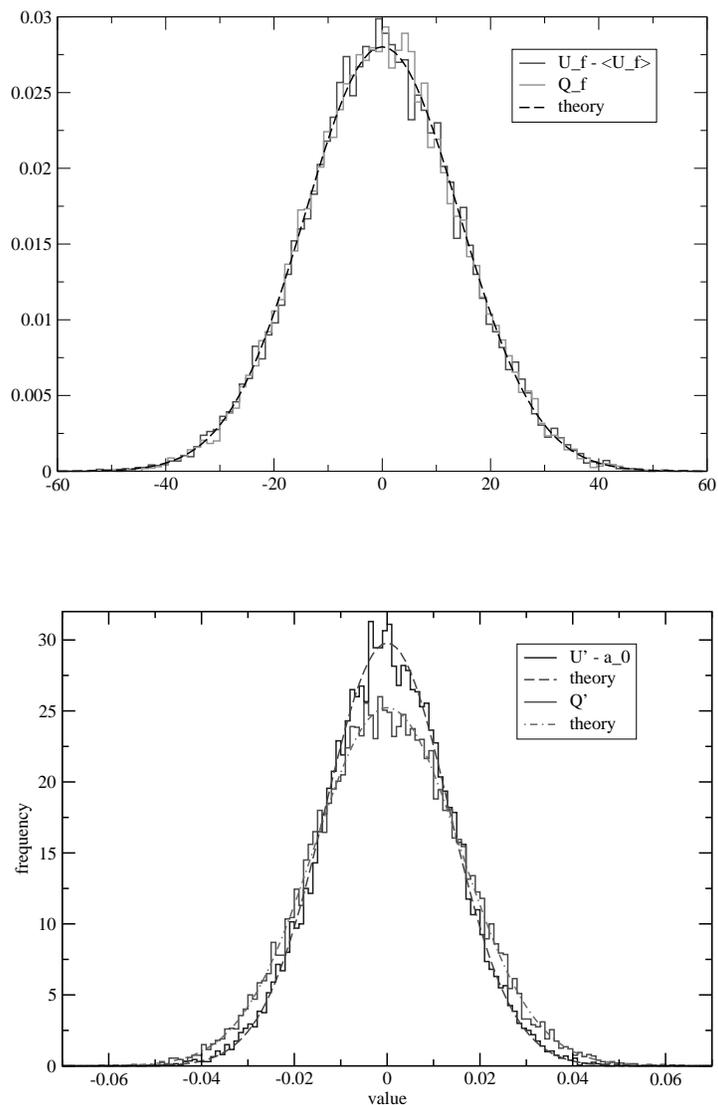

\centerline{\includegraphics[angle=0,scale=.4]{fig2topdistributionUQ.eps}}
\vspace {0.6in}
\centerline{\includegraphics[angle=0,scale=.4]{fig2bottomdistributionUQprime.eps}}
\caption{(top) Marginal distributions of $U_{f}$ and $Q_{f}$  for $20,000$ simulations with $<N>=4000$ for a polarization $a_{0} = 3/4 $ along the $U_{f}$ axis. The theoretical distributions are the same. (bottom) Marginal distributions of $U'$ and $Q'$ for $20,000$ simulations with $<N>=8000$  for a polarization $a_{0} = 3/4 $ along the $U_{f}$ axis. Here the ordinate is the number of simulations in a bin. The curves are the predictions for independent normal distributions. \label{Figb}}
\end{figure}
\clearpage
\begin{figure}
\vspace{-0.2in}
\centerline{\includegraphics[angle=0,scale=.55]{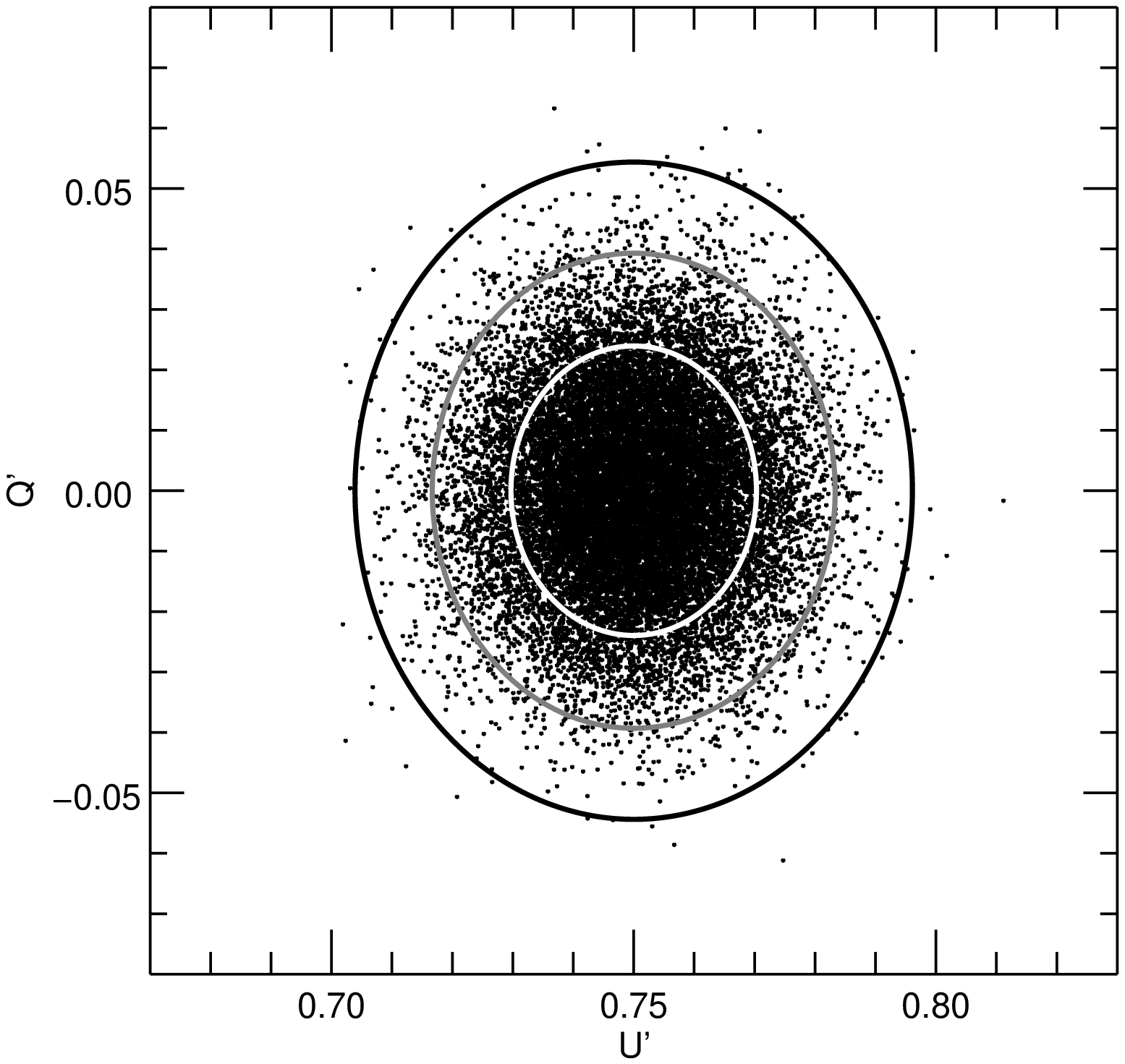}}
\vspace{-0.2in}
\centerline{\includegraphics[angle=0,scale=0.55]{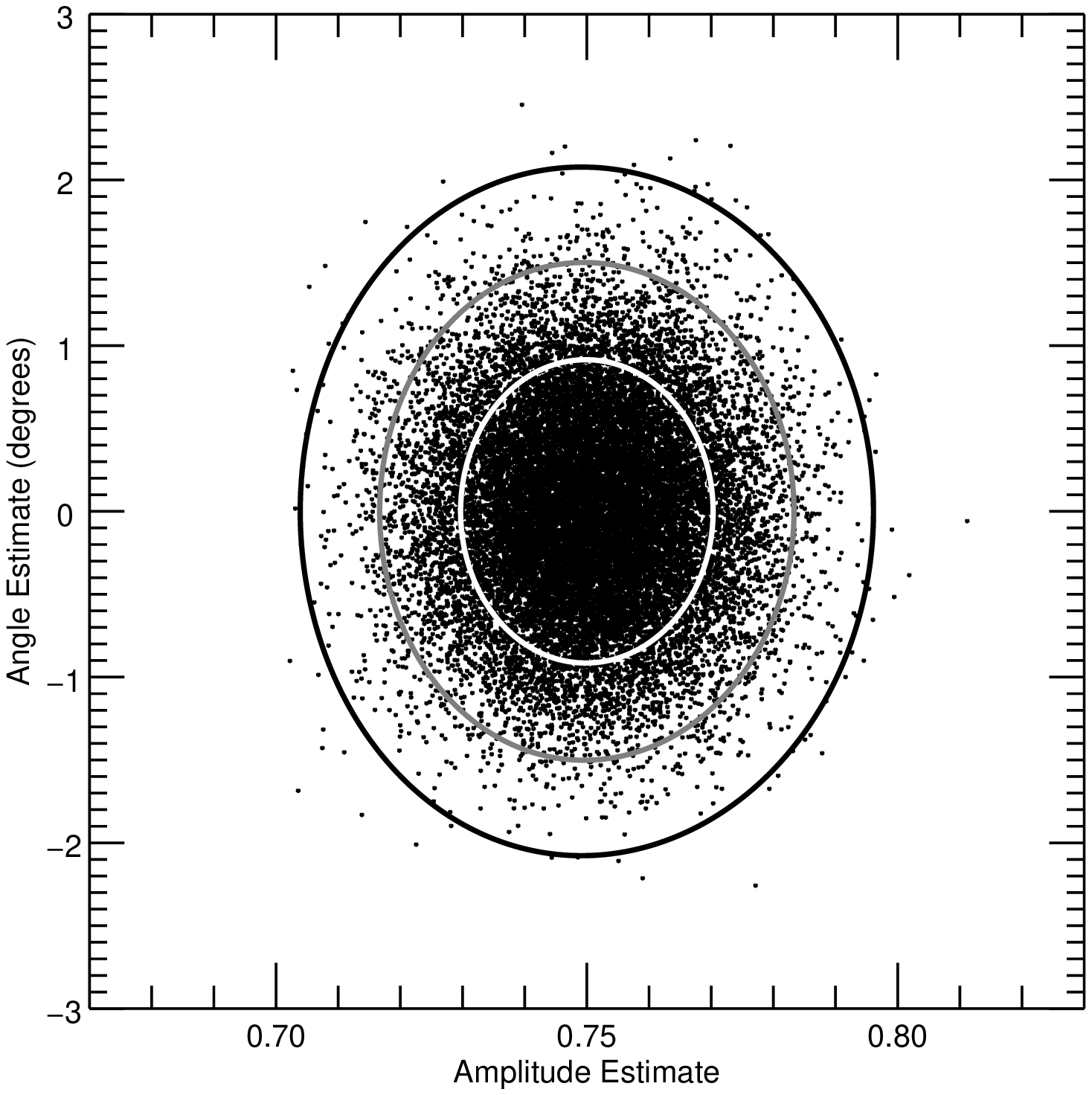}}
\vspace{-0.3in}
\caption{(top)Distribution of $U'$ and $Q'$  for $20,000$ simulations (the same as for Fig 2, bottom) with
$<N>=8000,  a_{0}=3/4,$ and $\phi_{0} = 0$. Contours of the predicted 1, 2, and 3 sigma levels (68.27, 95.45, and 99.73 \%) are superposed.  13597, 19010, and 19935 simulations fell within those contours, in comparison to 13654, 19090, and 19946 expected. The semi-major axes are along the $Q'$ axis and the semi-minor axes along the $U'$ axis, centered on $a_{0},0$. (bottom) The corresponding $a_{e} $ and $\phi_{e}$ with their theoretical contours.
\label{Figc}}
\end{figure}
\clearpage
\begin{figure}
\centerline{\includegraphics[angle=0,scale=0.6]{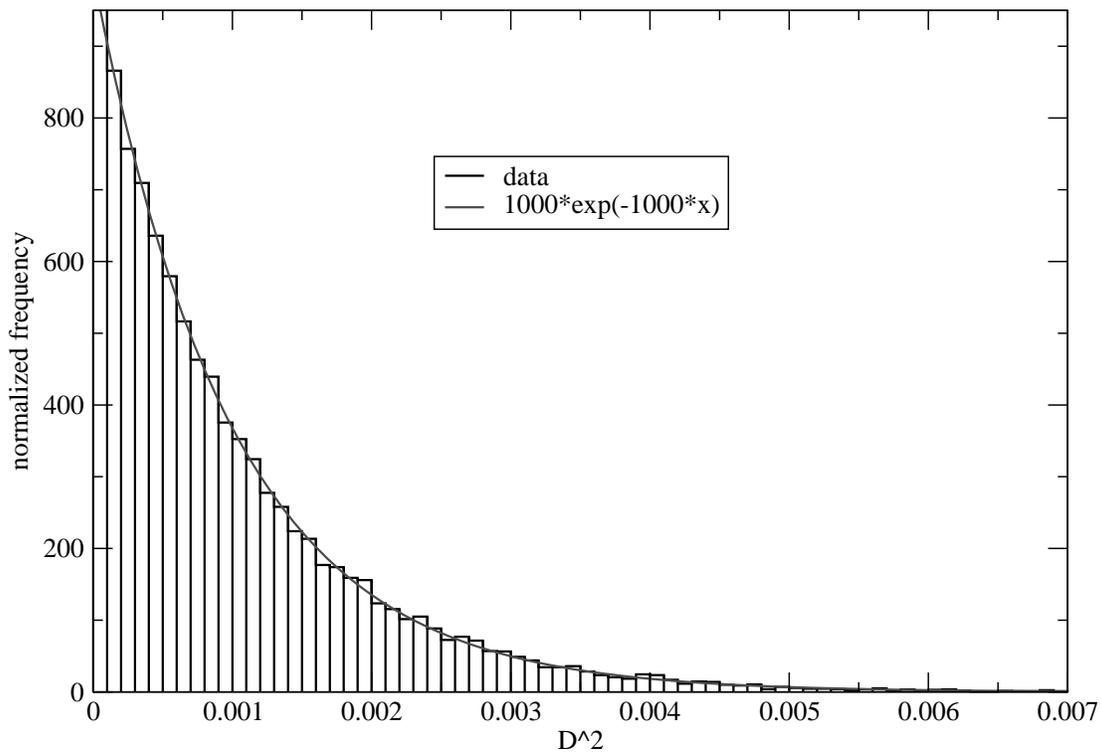}}
\caption{ Distribution in $D^{2}$ for a simulation with $a_{0}=3/4$ and $<N>=4000$, for $N_{sample} = 20,000$. The number of samples in a $D^{2}$ increment $\Delta D^{2} = 10^{-4}$, is plotted, normalized by $N_{sample} \Delta D^{2} (=2)$, together with the expected values. For this case $1/2\sigma'^{2} = 1000$. The probability for the measured polarization to lie outside of $D^{2}$ matches the predicted exponential. \label{Figd}}
\end{figure}
\clearpage
\begin{figure}
\vspace {-.2in}
\centerline{\includegraphics[angle=0, scale=0.55]{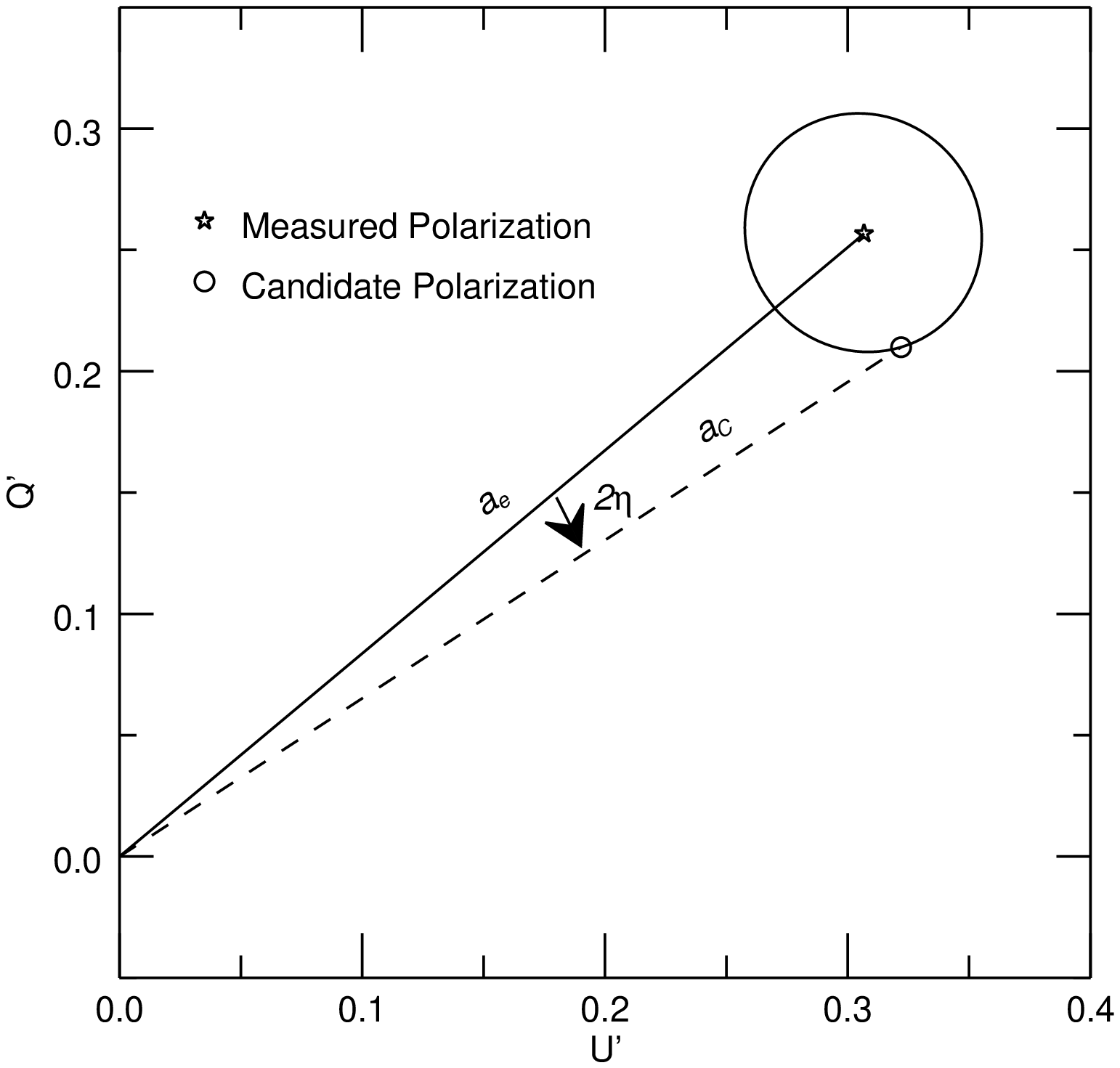} }
\vspace{-0.2in}
\centerline{\includegraphics[angle=0,scale=0.55]{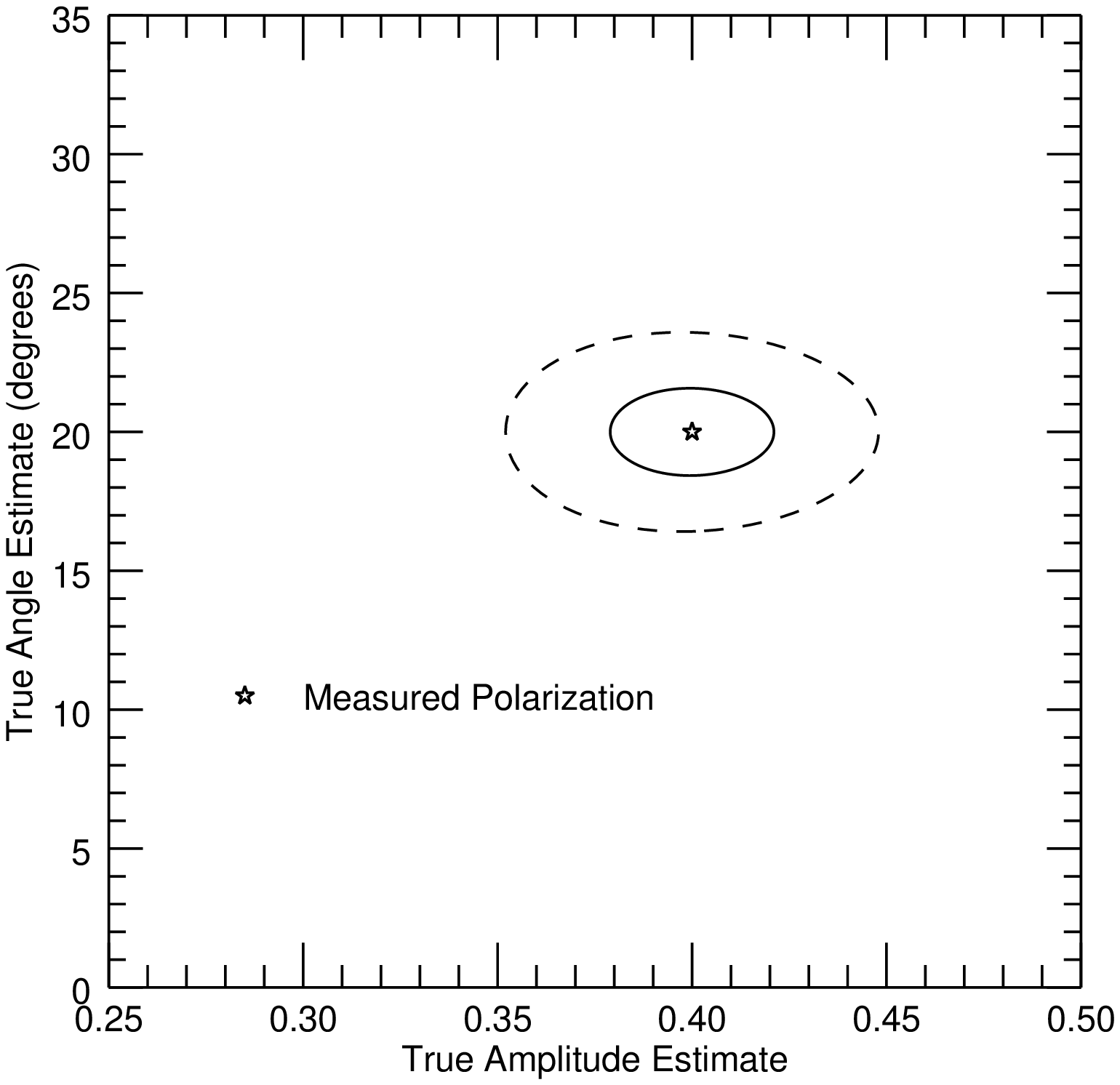} } 
\vspace{-0.3in}
\caption{Examples of contours for the candidate true polarization quantities for a measurement of $a_{e}=0.40$ and $\phi_{e}=20$ degrees. (top) Contour for 68.3 \%  (theoretical) confidence in $U' Q'$ space for the true polarization, for $D=0.05$. The dashed line indicates a candidate true polarization at distance $a_{C}$ from the origin and angle $2\eta$ from the measured polarization. (bottom) Contours (solid and dashed lines, respectively) of 1 sigma (68.3 \%) and 3 sigma (99.7 \%) for the true amplitude and angle, for $N=9500 $ events. \label{Fige}}
\end{figure}


\begin{thebibliography}{dummy} 
\bibitem[Clarke et al (1983)]{CSSB83} Clarke, D., Stewart, B. G., Schwarz, H. E., \& Brooks, A. 1983, A\&A, 126, 260
\bibitem[Elsner, O'Dell \& Weisskopf (2012)]{EOW12} Elsner, R. F., O'Dell, S. L., \& Weisskopf, M. C. 2012, Proc. SPIE, 8443, 84434N, astro-ph/1208.0610
\bibitem[Hamaker \& Bregman (1996)]{HB96} Hamaker, J. P., \& Bregman, J. D. 1996, A\&AS, 118, 161
\bibitem[Mardia \& Jupp (1999)]{MJ99} Mardia, K. V., \& Jupp, P. E. 1999, Directional Statistics (New York: Wiley), 45
\bibitem[Markwardt(2009)]{CM09} Markwardt, C.B. 2009, in ASP Conf. Ser. 411, Astronomical Data Analysis Software and Systems XVIII, ed. D.A. Bohlender, D. Durand, \& P. Dowler (San Francisco, CA:ASP), 251, astro-ph/0902.2850
\bibitem[Rice (1945)]{R45} Rice, S.O. 1945, Bell System Tech. J., 24, 46
\bibitem[Serkowski(1962)]{Ser62} Serkowski, K. 1962, AdA\&A, 1, 289
\bibitem[Simmons \& Stewart(1985)]{SS85} Simmons, J. F. L., \& Stewart, B. G. 1985, A\&A, 142, 100
\bibitem[Strohmayer \& Kallman(2013)]{SK13} Strohmayer, T. E., \& Kallman, T. R. 2013, \apj, 773, 103 (Paper I), astro-ph/1306.3885
\bibitem[Vaillancourt(2006)]{Vai06} Vaillancourt, J. E. 2006, \pasp, 118, 1340, astro-ph/0603010
\bibitem[Wardle \& Kronberg(1974)]{WK74} Wardle, J. F. C., \& Kronberg, P. P. 1974, \apj, 194, 249
\bibitem[Weisskopf et al (2009)]{WEK09} Weisskopf, M. C., Elsner, R. F., Kaspi, V. M., et al. 2009, in Neutron Stars and Pulsars (Astrophysics and Space Science Library, Vol. 357; berlin: Springer), 589
\bibitem[Weisskopf, Elsner, \& O'Dell(2010)]{WEO10} Weisskopf, M. C., Elsner, R. F., \& O'Dell, S. L. 2010, Proc. SPIE, 7732, 11, astro-ph/1006.3711
\bibitem[Yfontis \& Borgman (1982)]{YB82} Yfontis, E. A., \& Borgman, L. E. 1982, CoSTM, 11, 1695
\bibitem[Zelin \& Severo (1964)] {Mars64} Zelen, M. \& Severo, N. C. 1964, in  Handbook of Mathematical Functions, ed. M. Abramowitz \& I. A. Stegun (Washington, DC: Government Printing Office), 925

\end{thebibliography}
 \end{document}